# Interaction of highly nonlinear solitary waves with linear elastic media

Jinkyu Yang[1], Claudio Silvestro[2], Devvrath Khatri[1], Luigi De Nardo[2], and Chiara Daraio[1]*

[1] *Graduate Aerospace Laboratories (GALCIT), California Inistitute of Technology*
*Pasadena, CA 91125, USA*
[2] *Dipartimento di Chimica, Materiali e Ingegneria Chimica "G. Natta", Politecnico di Milano*
*Milano 20133, Italia*

We study the interaction of highly nonlinear solitary waves in granular crystals, with an adjacent linear elastic medium. We investigate the effects of interface dynamics on the reflection of incident waves and on the formation of primary and secondary reflected waves. Experimental tests are performed to correlate the linear medium geometry, materials, and mass with the formation and propagation of the reflected waves. We compare the experimental results with theoretical analysis based on the long-wavelength approximation and with numerical predictions obtained from discrete particle models. Studying variations of the reflected wave's velocity and amplitude, we describe how the propagation of primary and secondary reflected waves responds sensitively to the states of the adjacent linear media. Experimental results are found to be in agreement with the theoretical analysis and numerical simulation. This preliminary study establishes the foundation for utilizing reflected solitary waves as novel information carriers in nondestructive evaluation of elastic material systems.

## I. INTRODUCTION

Wave propagation and localization at the interface of different physical media has aroused great interest in diverse research areas such as solid-state physics [1], optics [2-3], and acoustic [4-6]. In particular, the study of interfaces between linear and nonlinear optical media has allowed the observation of interesting spatially localized phenomena known as optical Tamm states [7]. In linear lattice structures, acoustic localization has been reported in association with the boundary conditions [4] and with the presence of local defects in an otherwise periodical system [5].

Recently, one-dimensional (1D) granular media composed of contacting elastic particles (granular crystals) have been widely employed in the study of wave propagation [8-10] and acoustic vibrations [11-12]. It has been shown that the dynamic response of these chains of particles can encompass linear, weakly nonlinear, and strongly nonlinear regimes with highly scalable and tunable properties. In the strongly nonlinear regime, a granular chain supports the formation and propagation of highly nonlinear solitary waves (HNSWs) [8-9]. Unlike harmonic oscillatory waves in linear elastic media, HNSWs are lumps of energy that present unique scattering and superposition responses. They are characterized by a compactly-supported shape and extremely slow propagation velocity in comparison to the sound speed of the material that composes the particles in the chain [9, 13].

Solitary wave reflection and localization at a rigid boundary have been previously surveyed [14], but limited work has been done on the study of solitary waves interacting with linear elastic interfaces. Job *et al.* investigated the collision of a single solitary wave with elastic media of various hardness (herein referred to as the "wall"), and reported different force profiles originating from the interactions with such walls [15]. Falcon *et al.* studied the impact of a column of beads on a fixed wall focusing on the bouncing behaviors of the chain [16]. The decomposition of incident solitary waves has been reported under the influence of large-mass granular impurities [17-18] and heterogeneous granular chains [19].

In this study we systematically investigate the reflection of highly nonlinear solitary waves interacting with different linear elastic media. The primary focus is on analyzing the formation mechanisms and attenuation properties of the reflected waves as a function of the materials and geometry of the adjacent linear media. It was reported earlier that primary and secondary reflected solitary waves can form in granular media after a head-on collision between two solitary waves [14], or after the collision of an incoming solitary wave with a wall [15]. In this work, we quantitatively characterize the formation and propagation of secondary reflected solitary waves after the incoming wave interacts with various adjacent linear media. We describe delayed formation of the reflected wave as a function of the particles dynamics in the vicinity of the nonlinear and linear interface. Experimental tests are performed using a broad range of materials and geometries for the wall. We compare experimental results with theoretical results derived from the long wavelength approximation and with numerical results obtained from a discrete particles model based on Hertzian interactions. We observe that the reflected solitary waves are sensitive not only to the material properties of the immediately adjacent medium, but also to the properties and geometry of the underlying layers. The information-conveying characteristics of the reflected solitary waves make highly nonlinear granular chains very attractive for nondestructive evaluation of uniform or composite structures.

The rest of the manuscript is structured as follows: First, we describe the experimental setup in Section II. We then introduce a numerical model to explain the coupling between nonlinear and linear media in Section III. Section IV describes theoretical analysis of the particle dynamics at the interface. Section V describes a comparison between analytical, numerical, and experimental results. Lastly, in Section VI, we conclude the manuscript with summary and possible future work.

## II. EXPERIMENTAL SETUP

The nonlinear granular medium studied in this work consisted of a vertical chain of 20 stainless steel spheres (McMaster 440C with 4.76 mm radius and material properties listed in Table 1). The spheres were constrained by four steel rods coated by Teflon-tape to reduce friction [Fig. 1]. The impact of a spherical striker, identical to the spheres composing the chain, was used to generate a single solitary wave in the chain [9]. A DC-powered linear solenoid released the striker from a drop height of 1 cm to accurately control the striker velocity. This release method allowed us to obtain highly reproducible impacts, with only a 0.45% standard deviation in the velocity distribution. We limited our work to the study of 1 cm drop height in order to exclude the possible onset of plasticity at or around the contact region [20]. A high-speed camera (Vision Research Phantom V12) was employed to measure the actual impact and rebound velocity of the striker to characterize energy losses.

The propagation of solitary waves in the chain was recorded using two instrumented particles with calibrated piezo sensors. To fabricate an instrumented particle a thin layer of lead zirconate titanate ceramics (APC-850 PZT with 4.75 mm radius and 0.50 mm thickness) was embedded between two spherical caps as illustrated in the inset of Fig. 1. Particular care was taken to ensure that the total mass of the instrumented particle was equal to the mass of the regular beads in the chain. The piezo elements were electrically insulated by Kapton film (McMaster, low-static polyimide with 66.0 μm thickness) to prevent charge leakage to the neighboring elements. All the components were assembled together using epoxy adhesives. Custom microminiature wiring was soldered on the silver coated electrodes of the piezo ceramics to allow connection to a Tektronix 2024 oscilloscope for signal acquisition. The instrumented particles were positioned in the 7th and 16th positions from the top of the chain. The voltage-to-force conversion factors were obtained based on conservation of momentum as described in [13].

We tested 4 different sets of cylinders in order to simulate various states of linear media. In each test, the bottom of the sample was firmly fixed to a massive V-block by steel adaptors and clamps with fastening screws to impose fixed boundary conditions [Fig. 1]. First we tested uniform cylinders made of various materials to assess the effect of their mechanical properties on the solitary wave reflection. The materials tested ranged from soft polymers to hard metals, with their properties listed in Table 1. The cylindrical samples were 76.2 mm tall with radius 19.1 mm, 4 times larger than the radius of the spheres in the chain. By using samples with a large cross-sectional area, we can reduce the boundary condition used in the theoretical analysis to a semi-infinite wall [21-22].

The second set of cylindrical samples was prepared to examine the effect of the cylinder's geometry. In this case, we tested slender stainless steel cylinders (9.525 mm radius, twice the radius of beads used in granular chain) and took into account their dynamics in relation to their longitudinal dimensions using a 1D approximation. We examined 14 samples with various heights ranging from 6.35 mm to 610 mm. The cylinder centerlines were aligned with the axis of the granular chain to prevent the generation of flexural waves.

The third and fourth sets of samples consisted of layered cylindrical media. We assembled double-layered composite media out of two distinct materials [Fig. 2(b)]. In this setup, cylinders of 440C stainless steel were glued on top of polytetrafluoroethylene (PTFE) rods with epoxy adhesive. The radii of both the stainless steel and PTFE cylinders were 9.525 mm. We first tested different heights of the stainless steel cylinders (from 6.35 mm to 101.6 mm) positioned on the top of a 25.4 mm-tall PTFE cylinder. The use of steel cylinders of different heights allowed evaluating the effect of the upper layer's inertia on the formation of reflected waves at the interface. We then tested different heights of the PTFE lower cylinders (from 9.52 mm to 152.4 mm), while keeping constant the geometry of the stainless steel part with a height of 6.35 mm. Using these specimens, we investigated the effect of the lower material's stiffness on the solitary wave interaction.

Table 1. Material properties of polymer and metallic specimens: The reported values are standard specifications [23] except the Young's moduli of polymers, which are extrapolated from Hugoniot relationship [24].

| Material | Density [kg/m$^3$] | Young's modulus [GPa] | Poisson's ratio |
|---|---|---|---|
| Stainless steel AISI type 440C | 7800 | 200 | 0.28 |
| Copper | 8960 | 115 | 0.35 |
| Brass (360) | 8550 | 103 | 0.34 |
| Aluminum 6061-T6 | 2693 | 68.9 | 0.33 |
| Nylon (Polyamide) | 1140 | 6.52 | 0.40 |
| Acrylic (Polymethylmethacrylate) | 1186 | 4.75 | 0.35 |
| Polycarbonate | 1196 | 3.75 | 0.35 |
| PTFE (Polytetrafluoroethylene) | 2151 | 1.53 | 0.46 |

### III. NUMERICAL MODEL

To evaluate the wave dynamics in the granular chain we used a 1D discrete particle model [9]. In this approach, we assumed that the particles interaction is restricted to the axial direction and that the transit times of the solitary waves in the granular media are much longer than the oscillation period of elastic waves within the particles. Under these 1D assumptions we can express the equation of motion of monodispersed spherical particles using a modified Hertzian model that includes dissipative terms [25]:

$$m\ddot{u}_n = \left(A_n \delta_n^{3/2} - A_{n+1} \delta_{n+1}^{3/2}\right) + \left(\gamma_n \dot{\delta}_n - \gamma_{n+1} \dot{\delta}_{n+1}\right) + F \qquad n \in \{1,\ldots,N\} \qquad (1)$$

where

$$A_n \equiv \begin{cases} A\big|_c = \dfrac{E\sqrt{2R}}{3(1-v^2)} & n \in \{1,\ldots,N\} \\ A\big|_w = \dfrac{4\sqrt{R}}{3}\left(\dfrac{1-v^2}{E}+\dfrac{1-v_w^2}{E_w}\right)^{-1} & n = N+1 \end{cases}$$

$$\gamma_n \equiv \begin{cases} \gamma\big|_c & n \in \{1,\ldots,N\} \\ \gamma\big|_w & n = N+1 \end{cases}$$

$$\delta_n \equiv [u_{n-1}-u_n]_+ \quad n \in \{1,\ldots,N+1\}.$$

Here $R$ is the radius of the bead, and $u_n$ is the coordinate of $n$-th bead's center from its equilibrium position. We represent the striker bead with index $n = 0$, and the displacement at the nonlinear and linear media interface is denoted by $u_{N+1}$ [Fig. 2(a)]. The bracket $[s]_+$ takes only positive values and is equal to 0 if $s \leq 0$. The subscripts $|c$ and $|w$ refer to the chain and wall respectively. Here $F$ is a body force applied to the bead – gravity in this study, and $E$, $m$, and $v$ represent Young's modulus, mass, and Poisson's ratio of the granular elements.

It should be noticed that $A\big|_w$, the stiffness value between the last bead and the bounding wall, is different from $A\big|_c$, the stiffness between two beads, due to the sphere-wall contact configuration. This sphere-wall stiffness value depends on the mechanical properties of the wall, which are characterized by Young's modulus $E_w$ and Poisson's ratio $v_w$ of the wall. We define the critical Young's modulus of the wall, when the wall stiffness becomes identical to the chain stiffness ($A\big|_c = A\big|_w$). Mathematically, the critical Young's modulus is expressed as:

$$E_w\big|_{critical} = \frac{E}{2\sqrt{2}-1}\left(\frac{1-v_w^2}{1-v^2}\right), \tag{2}$$

based on the chain and wall stiffness definition in Eq. (1). If the linear elastic cylinder and the chain particles exhibit similar Poisson's ratios, this relationship can be simplified to $E_w\big|_{critical} \approx E/(2\sqrt{2}-1)$, where the critical elastic modulus of the wall is approximately 55% of the particle material's modulus. We later discuss the physical meaning of this critical Young's modulus.

Previous studies focused only on the energy dissipation along the chain [15, 25]. It is of particular importance for us to account also for restitutional losses against the wall for the accurate prediction of the wave reflection behavior at the interfaces. We found that a linear dissipation model [26] yields satisfactory results to describe the damping phenomena in short chains of granular particles. The energy losses along the chain are represented by the chain dissipation coefficient $\gamma\big|_c$. The damping at the interface is characterized by the wall dissipation coefficient $\gamma\big|_w$, which has a strong dependence on the wall materials. The values of all damping coefficients were extrapolated from experiments [see details in the Appendix].

We used the time delay system model [27] to simulate the wave propagation within the linear elastic medium. In this approach, we first define the dimensionless displacement field of the linear medium $u_w^*(\xi,\tau) \equiv u_w(x,t)/L$ with $\xi = \dfrac{x}{L}$ being the dimensionless position and $\tau = \dfrac{ct}{L}$ being a time parameter [Fig. 2(a)]. Here $x$ and $L$ are the position and the length of the linear medium, and $c$ and $t$ are the longitudinal wave speed and time. The displacement field can be expressed as $u_w^*(\xi,\tau) = f(\tau-\xi) + g(\tau+\xi)$ with the two real functions *f* and *g* representing forward and backward waves to satisfy the D'Alembert's general solution for the longitudinal wave equation [27].

The wave propagation solution in the linear medium is subjected to the boundary and initial conditions. The force exerted on the medium by the last bead is governed by the Hertzian interaction including the dissipative force, which is in equilibrium with the elastic repulsion by the linear medium:

$$-E_w S \, \partial u_w^* / \partial \xi \big|_{\xi=0} = A\big|_w \delta_{N+1}^{3/2} + \gamma\big|_w \dot{\delta}_{N+1}, \tag{3}$$

where *S* is the cross-sectional area of the linear medium. Using $\partial u_w^*/\partial \xi\big|_{\xi=0} = -f'(\tau) + g'(\tau)$ and $u_{N+1} = L[f(\tau) + g(\tau)]$, Eq. (3) can be expressed as:

$$E_w S[f'(\tau) - g'(\tau)] = A\big|_w \delta_{N+1}^{3/2} + \gamma\big|_w \dot{\delta}_{N+1},$$
$$\delta_{N+1} = [u_N - L\{f(\tau) + g(\tau)\}]_+. \tag{4}$$

It is notable that Eq. (4) includes the displacement terms of both the granular chain and the linear medium, linking the dynamics of the nonlinear and linear medium.

We now apply the rest of boundary and initial conditions to the linear elastic medium model. Given that the base of the cylinder is fixed (at $x = L$), we obtain the Dirichlet boundary condition:

$$f(\tau-1) + g(\tau+1) = 0. \tag{5}$$

The initial condition is that the linear medium remains undisturbed until time $t_0$, which denotes the moment that the solitary waves arrive in the end of the chain. This means that $f(\tau) = 0$ for $\tau \leq ct_0/L$ and $g(\tau) = 0$ for $\tau \leq 1 + ct_0/L$. As expressed in Eq. (5), the two functions *f* and *g* are essentially identical but shifted with a time delay term. For the numerical integration of such time delay systems, we used the DDE (delay differential equations) solver in MATLAB [27-28]. More complicated cases of multi-layered composite media [Fig. 2(b)] can be easily modeled by introducing additional pairs of real functions and the corresponding boundary conditions.

## IV. THEORETICAL ANALYSIS

It is well known that the energy carried by a solitary wave exists in a lumped form, mostly confined within the wave length of five particles in spherical chains [9]. During the impact on the wall, in particular, a majority of the propagating energy is retrieved as potential energy between the last bead and the bounding wall [15]. To analytically estimate the contact time of the last particle in the chain and the wall, we assumed that the collision process is fully elastic [29-30]. In this model the total energy initially carried by the striker is split between the Hertzian potential and kinetic energy of the last bead against the bounding wall. Mathematically:

$$\frac{1}{2} m V_s^2 \approx \frac{1}{2} m \left(\frac{du_N}{dt}\right)^2 + \frac{2}{5} A\big|_w u_N^{5/2}, \tag{6}$$

where $V_s$ is the striker velocity. Integrating this differential equation over the period of interaction, the contact time $T_c$ is obtained as:

$$T_c \approx 3.218 \ m^{2/5} V_s^{-1/5} A\big|_w^{-2/5}. \tag{7}$$

This equation shows the dependence of the contact time on the cylinder's mechanical properties as represented by $A\big|_w$ [see Eq. (1)].

We proceed to calculate the travelling time of the solitary waves in the granular chain between the instrumented bead and the wall. The double transit time $T_t$ of the solitary waves along the chain can be expressed as $T_t = d/V_i + d/V_r \approx 2d/V_i$. Here, $V_i$ and $V_r$ are the incident and reflected solitary wave velocity, and $d$ is the distance between the centers of the last bead and the instrumented bead. In this approximation we assumed that the reflected solitary wave velocity is almost the same as the incident velocity. We found that the difference between the incident and the reflected solitary wave velocities is not large even in the case of solitary waves interacting with a "soft" wall, where the reflected waves are highly attenuated. This is because the effect of the force amplitude on the solitary wave velocity is extremely weak; $V_i \propto F_m^{1/6}$, where $F_m$ is the maximum force between the beads [9]. Furthermore, the error induced by the velocity discrepancy is relatively small, compared to the elongated contact time $T_c$ in the soft-wall impact.

Using Nesterenko's long-wavelength approximation [9], we can analytically derive the solitary wave propagation speed and thus, the wave travelling time $T_t$ in an uncompressed 1D monodispersed chain. According to [9], the speed of the incident solitary wave $V_i$ can be expressed in terms of bead velocity $\upsilon$ and chain stiffness $A\big|_c$:

$$\begin{aligned} V_i &= (16/25)^{1/5} (2R)\, (\upsilon\, A\big|_c^2 / m^2)^{1/5} \\ &\approx 1.829\, R\, (\upsilon\, A\big|_c^2 / m^2)^{1/5}. \end{aligned} \tag{8}$$

Under the excitation of a single solitary wave via the same-mass striker, Chatterjee [31] numerically found $\upsilon \approx 0.682 V_s$. Using $d \approx N \times 2R$, where $N$ is the number of beads between the sensor and the wall, the solitary wave travelling time $T_t$ in the granular chain becomes:

$$T_t = 4NR/V_i \approx 2.361\, N\, m^{2/5} V_s^{-1/5} A\big|_c^{2/5}. \tag{9}$$

We refer to the transit time between the incident and the reflected solitary waves as the time of flight (*TOF*). Hence, the *TOF* of the first arriving reflected wave can be expressed as $TOF \equiv T_c + T_t$. In this manuscript, we refer to the first reflected solitary waves as primary solitary waves (PSW). By combining the analytical formula in Eqs. (7) and (9), the *TOF* of the primary solitary wave is obtained as:

$$TOF\big|_{PSW} \approx (3.218\, A\big|_w^{-2/5} + 2.361 N A\big|_c^{-2/5})(m^2/V_s)^{1/5}. \tag{10}$$

The *TOF* values of the solitary waves in double-layered composite media can be also acquired based on our finding in the uniform media. Under the approximation of uncoupling with the rest of the chain, we can establish a simple elastic collision model between the end sphere and the top layer of the medium [32]. Based on momentum and energy conservation, the ratio of the end-particle's reflection velocity $\upsilon_r$ to the incident velocity $\upsilon_i$ can be approximated as $\dfrac{\upsilon_r}{\upsilon_i} = \dfrac{m - M_u}{m + M_u}$, where $m$ is the mass of the last bead, and $M_u$ is the mass of the upper layer. In the case that the upper layer is heavier than the particle mass ($M_u > m$), the end-particle always rebounds from the adjacent composite medium, triggering the formation of the primary solitary waves.

Based on Eq. (8), we can express the ratio of reflected solitary wave velocity to that of the incident wave in terms of particle velocities: $\frac{V_r}{V_i} = \left(\frac{v_r}{v_i}\right)^{1/5}$. Thus, the modified travelling time $T_t'$ of the solitary wave after the consideration of differing reflection speed becomes:

$$T_t' = \frac{d}{V_i}\left(1 + \frac{V_i}{V_r}\right) = \frac{T_t}{2}\left[1 + \left(\frac{M_u - m}{M_u + m}\right)^{-1/5}\right]. \tag{11}$$

Using Eqs. (7) and (11), the *TOF* of the primary solitary waves in the composite medium is:

$$\begin{aligned} TOF|_{PSW} &\approx T_c + \frac{T_t}{2}\left[1 + \left(\frac{M_u - m}{M_u + m}\right)^{-1/5}\right] \\ &= \left[3.218 A|_w^{-2/5} + 1.181 N A|_c^{-2/5}\left\{1 + \left(\frac{M_u - m}{M_u + m}\right)^{-1/5}\right\}\right](m^2/V_s)^{1/5}. \end{aligned} \tag{12}$$

Note that this equation approaches Eq. (10) in the limit of a semi-infinite wall assumption, where the mass of the upper media becomes infinite ($M_u \to \infty$).

The generation and propagation of the subsequent reflected waves can be also obtained in the double-layered medium. In this manuscript, the subsequent solitary wave reflected from the interface is referred to as secondary solitary wave (SSW). Based on the aforementioned simple collision model, we acquire the velocity of the upper layer $v_u = \frac{2m}{m + M_u}v_i$ after the impact by the end-particle. We simplify the displacement of the upper medium's center to $\frac{y_{N+1} + y_{N+2}}{2} \approx y_{N+2}$ [see Fig. 2(b)], since the upper stainless steel layer is much harder than the lower PTFE medium. Hence, the equation of motion of the upper layer during the collision with the lower layer becomes:

$$\frac{1}{2}M_u v_u^2 \approx \frac{1}{2}M_u\left(\frac{du_{N+2}}{dt}\right)^2 + \frac{1}{2}k_d u_{N+2}^2. \tag{13}$$

Under the linear elastic assumption, the stiffness of the lower medium can be expressed as $k_d \equiv E_d A/L_d$, where $E_d$ is the elastic modulus of the lower medium and $L_d$ is its length. After integrating Eq. (13), we obtained the contact time between the upper and lower linear media as $T_l = \pi\sqrt{\frac{M_u}{k_d}}$. The *TOF* of the secondary solitary waves can be now expressed as a sum of the *TOF* of the primary solitary wave in Eq. (12) and the linear medium contact delay $T_l$:

$$\begin{aligned} TOF|_{SSW} &\approx TOF|_{PSW} + T_l \\ &= \left[3.218 A|_w^{-2/5} + 1.181 N A|_c^{-2/5}\left\{1 + \left(\frac{M_u - m}{M_u + m}\right)^{-1/5}\right\}\right](m^2/V_s)^{1/5} + \pi\sqrt{\frac{M_u}{k_d}}. \end{aligned} \tag{14}$$

Here we neglected the minute difference in transit time caused by the attenuation of the secondary solitary waves. The analytical *TOF* values derived from Eqs. (12) and (14) are verified in the next section using numerical and experimental approaches.

## V. RESULTS AND DISCUSSION

### A. Effects of Young's modulus

We first assessed the effect of the elasticity of the uniform linear media on the formation and propagation of reflected solitary waves. Figure 3 reports their experimental and numerical results of solitary wave interaction with the "soft" PTFE ($E = 1.53$ GPa) and the "hard" stainless steel ($E = 200$ GPa). The first impulse corresponds to the incident solitary wave generated by the striker impact, and the subsequent impulses represent the solitary waves reflected against the linear media, all detected on the 7th particle from the impacted end (14th particle away from the interface). The striker impact was identical in the two cases, leading to the formation and propagation of similar incident solitary waves. However, the reflected force profiles present a clear difference between the two cases. Most notably, the interaction with the soft interface results in a delayed formation of the primary reflected solitary wave (PSW) and in the generation of secondary reflected solitary waves (SSW) [Fig. 3]. Such unique behaviors are more obvious from the surface plot in Fig. 4, which reports numerically the temporal formation of primary and secondary reflected solitary waves over a range of the wall elastic modulus. It is evident that as the elasticity of the bounding media decreases, the *TOF* of the reflected wave increases. We also detect the formation of the secondary solitary waves under the soft-wall interactions. The formation of secondary solitary waves at the interface with a bounding wall was first reported in [15] but never studied systematically.

To elucidate the SSW generation mechanism, we numerically investigated the displacement profiles of the granular particles, before and after the wave reaches the "hard" stainless steel medium [Fig. 5(a)] and the "soft" PTFE medium [Fig. 5(b)]. The displacement curves differ significantly between the two cases in the post-reflection period. In particular, the last bead in the chain, denoted by the *N*-th particle, exhibits a remarkably different behavior. According to our numerical simulations, the PTFE medium allows four times deeper penetration of the last bead into the wall than the stainless steel wall, caused by the discrepancy in the wall stiffness between the stainless steel and PTFE materials; $A|_w$ for the stainless steel wall is 57 times larger than that of the PTFE wall.

The penetration process into the "soft" material can result in the loss of contact of the last bead with the rest of the chain. In this case, the last bead experiences a significant delay time until it bounces back under the resistance of the elastic medium. At 450 μs marked by the circle in Fig. 5(b), the rebounding bead collides with the rest of the chain. This first collision generates the formation of the primary reflected solitary wave. After approximately 0.1-ms delay, we observed that the last bead and the rest of the chain undergo the second impact due to the wall elasticity, corresponding to the 561-μs moment marked by the diamond in Fig. 5(b). This second encounter triggers the formation of the secondary solitary wave. After the second collision, subsequent minor impacts are followed producing small backscattered waves in the granular chain. In contrast to the multiple impacts in the "soft" wall case, the last particle experiences a single strong collision with the rest of the chain, when the incoming solitary wave interacts with a "hard" adjacent medium [Fig. 5(a)]. In this case no gap opening is created between the particles, and most energy is retrieved by the PSWs in the absence of noticeable SSWs.

We hypothesize that the secondary solitary waves occur when the last particle in the chain loses contact with the others. This gap opening between the last bead and the rest of the chain is guaranteed if the end-particle's penetration depth exceeds its neighboring bead's maximum displacement in a non- or weakly-compressed granular chain. The penetration depth of the last bead becomes identical to the maximum displacement of its neighboring bead in the condition $A|_c = A|_w$, which yields the critical

Young's modulus as defined in Eq. (2). From the numerical results in Fig. 4, we find that the secondary solitary waves become noticeable after this critical elastic modulus ($E \approx 100$ GPa).

We compared the numerical, analytical, and experimental results of the *TOF* and amplitudes of the reflected waves as a function of the Young's modulus of the linear media [Fig. 6]. To calculate the error bars in experiments we repeated five tests per specimen and computed the averages and standard deviations of their *TOF* and amplitude ratios. We observed that the *TOF* of the primary solitary wave is significantly longer in the "soft" polymer media than in the "hard" metallic specimens. Viewed from the instrumented bead in the chain, the average *TOF* of solitary waves against the PTFE cylinder is 0.616 ms, 45% longer than the 0.483-ms *TOF* in the case of the stainless steel cylinder. We found that the experimental results are in a good agreement with the numerical simulations and analytical results obtained from Eq. (10). We also compared the ratio of the primary reflected solitary waves' amplitudes to those of the incident solitary waves. As illustrated in Fig. 6(b), our experimental data indicate that the amplitude ratio values range from 42.1% for PTFE to 77.0% for stainless steel materials. We found that adjacent media with high stiffness generated large-amplitude reflected waves, whereas softer adjacent media produced substantially attenuated primary reflected waves. The general trend of the reflection ratios obtained numerically and experimentally in this study agrees well with the results reported by Job *et al.* [15].

Figures 6(c) and 6(d) report the time of flight and amplitude ratio of secondary reflected solitary waves as a function of the stiffness of the bounding media. The *TOF* curve of the secondary reflected solitary waves shows an analogous trend as that of the primary solitary waves with an approximately 0.1-ms delay. However, the ratio of force amplitudes of PSWs [Fig. 6(b)] and SSWs [Fig. 6(d)] show strikingly different patterns. The behavior of the SSW amplitudes resembles a Sigmoidal-shaped function. Contrary to the PSW behavior, the presence of a "hard" adjacent medium produces a smaller secondary solitary wave than that generated by a "soft" adjacent medium. This trend is in accordance with momentum conservation [15]. The plateau in the low elastic modulus range is probably due to the increased amount of restitutional energy losses when secondary solitary waves are generated under the interaction with a "soft" wall.

**B. Effects of uniform layer thickness**

Figure 7 shows the numerical and experimental results of *TOF* and amplitude ratios as a function of the cylinder heights. Interestingly, the reflected solitary waves did not show any significant dependence on the heights of the slender linear media. The distribution of *TOF* values is extremely regular in the 0.44 ~ 0.45 ms range [Fig. 7(a)]. The numerical results predicted slightly longer *TOF* values, but the discrepancy is minute. The values of the amplitude ratios also remain between 73% and 78% range [Fig. 7(b)]. This implies a consistently strong reflection of the solitary waves over a range of specimen lengths. It is also notable that no secondary reflected solitary wave was observed in the range of stainless steel cylinders' geometry despite the large variation in sample sizes tested.

To understand the observed behavior of the solitary wave reflection, we analyzed the wave dynamics in the vicinity of the contact point. When the incoming solitary wave arrives at the interface, the last bead in the chain starts to interact with the linear medium applying compressive force to its top surface. Under the 1D approximation, neglecting surface and flexural waves in the rod, a longitudinal wave propagates along the axial direction of the rod and reflects back from the opposite end. During $T_c$, the contact time of the last bead on the linear medium, the longitudinal pressure wave travels a distance equal to $cT_c$, where $c$ is the speed of the longitudinal wave in the elastic medium. In a long cylindrical member with length $L > cT_c/2$, this longitudinal pressure wave does not return to the contact interface during the contact time $T_c$. In this case a portion of the incident energy from the granular chain is lost at the interface in the form of leaked elastic waves into the rod. On the other hand, if $L < cT_c/2$, the injected energy is partially transferred back to the nonlinear granular chain during the solitary wave interaction with the cylinder. Using the analytical contact time in Eq. (7), we obtained the characteristic length

$cT_c/2$ = 101 mm for the given setup. We observed a reduction (~5%) in the amplitude ratio around this characteristic length in the numerical results [Fig. 7(b)]. However, it was difficult to capture such reduction in experiments due to the limited sensitivity of the experimental setup.

The negligible sensitivity to the height of the linear medium implies that the energy lost by elastic waves into the linear media is not significant. Previous studies have shown that only a small amount of energy is lost due to the wave propagation in the process of a sphere impact on the wall; In terms of restitution coefficient defined as the reflected velocity over the incident velocity, the loss is in the range of 0.5% to 3.0% [29-30]. Such a minor effect of energy loss is in agreement with our observations. It should be noted that the negligible sensitivity is valid only if the wall is much more rigid than the chain. If the cylindrical member is made of a "soft" material or if its cross-sectional area is extremely small, the incidence of solitary wave at the interface results in a considerable compression of the linear medium during the contact time. Consequently, the particle dynamics in the vicinity of the wall is inevitably affected by the linear medium stiffness. In Section D, we investigated such stiffness effect caused by the "soft" media's geometrical change using composite media. For the stainless steel cylinders considered in this Section, however, the maximum displacement of the rod tip remained less than 0.28 μm even for the longest sample. This is an order of magnitude smaller than the maximum bead displacement in the chain (8.2 μm) according to our numerical simulations. Hence, under the interaction with "stiff" cylinders, the wave dynamics at the nonlinear/linear interface is not sensitively affected by the cylinders' uniaxial compression.

**C. Effects of upper layer thickness in composite media**

Now we consider the interaction of solitary waves with composite media. Figure 8 shows the numerical and experimental results of the solitary wave decomposition as a function of $L_u$, the upper stainless steel layer's height. Based on the simple collision model in §IV, it is evident that the larger inertia of the upper layer results in stronger repulsion of the granular chain, and consequently creates a higher reflection ratio of the primary solitary wave. Accordingly, the time of flight of the reflected solitary waves is also shortened, because a stronger repulsion increases the propagation speed of the reflected solitary waves. As shown in Fig. 8(a), numerical simulation predicts that as the height of the upper media increases from 6.35 mm to 101.6 mm, the *TOF* of primary reflected solitary wave drops from 0.470 ms to 0.455 ms. This trend is confirmed by the experimental results within error margin. The analytical result is also shown based on simple elastic collision theory as expressed in Eq. (12). Despite a noticeable offset, the analytical curve corroborates the numerical and experimental results. We find that the variation in the time of flight is not significant, as expected from the weak influence of the upper-layer mass on the solitary wave velocity. On the other hand, the reflection ratios obtained experimentally show more drastic changes from 0.428 to 0.769, when the height of the upper layer increases 16 times from 6.35 mm to 101.6 mm [Fig. 8(b)].

The behavior of secondary reflected solitary waves reveals a different trend. When the height of the upper cylinder is increased from 6.35 mm to 101.6 mm, we observed that the experimental arrival time on the instrumented sensor increases by 32.5%, from 0.619 ms to 0.821 ms, [Fig. 8(c)]. Comparing this with the *TOF* curve of the primary solitary waves [Fig. 8(a)], we noticed that the secondary solitary waves showed opposing behaviors. We found that the numerical *TOF* curve lies below the experimental curve with discernible error. This discrepancy probably stems from the incapability of the numerical model to capture the viscoelastic effect inside the lower PTFE media. The estimated time of flight based on Eq. (14) is plotted in Fig. 8(c), and it is in excellent agreement with the experimental results. This again confirms the elongated delay time of the SSW formation resulted from the increased inertia of the upper layer. The comparison is shown up to the height of 25.4 mm, because the secondary solitary waves are no longer noticeable in experiments after this point [Fig. 8(d)]. As shown in Fig. 8(b) and (d), the reflection ratios of the primary and secondary solitary waves balance, as we expect from momentum and energy conservation.

**D. Effects of lower layer thickness in composite media**

We studied how the thickness of the lower layer of the composite medium affects the formation of reflected solitary waves. We compared numerical and experimental results of extracted *TOF* and amplitude reflection ratios in Fig. 9. Both the numerical and experimental *TOF* values varied between 0.46 ms and 0.47 ms, showing only 2.5% fluctuations over a range of lower media's length variation. Analytical calculations of *TOF* values based on Eq. (12) are also presented in Fig. 9(a), and they are in qualitative agreement with the numerical and experimental results. The analytical predictions consistently underestimated the *TOF* values as shown in both Figs. 8(a) and 9(a). This is most likely due to the presence of dissipation in experiments, which is not accounted for in the analytical model. The reflection ratio is distributed between 0.40 and 0.46 both numerically and experimentally. The relative insensitivity of the PSWs to the lower medium dimension confirms that the generation of the PSWs is governed by the upper media properties.

Figure 9(c) reports the formation of the secondary solitary waves in terms of the analytical, numerical, and experimental *TOF* values. Compared to the PSW, the secondary solitary waves exhibit enhanced sensitivity to the lower base's stiffness. As predicted by the elastic collision theory in §IV, we observe that the longer "soft" medium yields delayed formation of the SSWs. This is due to the compressive behavior of the lower layer during the contact of the granular chain to the composite medium. Such geometrical effect of the "soft" lower medium is in sharp contrast to the "hard" wall interaction, where the compression of the stainless steel rods was negligible due to the cylinder's high stiffness. The reflection ratios of the SSWs show no clear trend [see Fig. 9(d)]. Here, as elsewhere in this study, we observed that the empirical errors of the reflection ratios were larger than those of *TOF*. This is because the force measurements are more susceptible to experimental errors than the *TOF* measurement due to possible tilting of the sensor and bead misalignment.

**VI. CONCLUSION**

We performed investigation on the interaction of highly nonlinear solitary waves with linear elastic media. We found that the formation and propagation of reflected solitary waves from the interface is strongly governed by the elastic modulus and geometry of the adjacent linear medium. The mechanisms of the decomposition and attenuation of the reflected solitary waves were analyzed by examining the complex particle dynamics in the vicinity of the nonlinear and linear media interface. Using analytical, numerical and experimental approaches, we verified that the travel time and force magnitudes of the primary and secondary reflected waves were strongly associated with the material properties and geometrical configurations of uniform and composite media. It is notable that the reflected waves always retained a compact support without the presence of significant dispersion or attenuation. The robustness of the reflected solitary waves, as well as their sensitivity, makes them extremely useful as information carriers in nondestructive evaluation applications. We limited our work to the investigation of only the primary and secondary solitary waves, but further studies can be performed to relate the subsequent impulses of nonlinear waves with the properties of more complex structures, such as multi-layered or heterogeneous material systems.

**APPENDIX: ENERGY DISSIPATION IN COMBINED GRANULAR AND LINEAR MEDIA**

In this study, we considered the dissipation occurring both in the chain and at the wall interface. To assess the dissipation in the chain, we performed experiments using the identical setup as described in §II, except that the chain was compose of 28 particles instead of 20 to acquire a longer stream of attenuation trends. The gradual attenuation of propagating solitary waves was recorded by shifting positions of an instrumented bead to every even numbered particle site in the chain. We acquired five data sets and averaged them to obtain a force profile for each sensor. Figure 10(a) shows the measured force profiles. The high spike in the middle is due to the direct interaction of the last particle with the wall [15].

We used the least square fitting method [33] to find the optimized chain dissipation coefficient $\gamma|_c$ that best matches with the experimental data. To exclude the restitutional dissipation effect, we considered only the incident solitary waves. The residual $R_i(\gamma)$ for the $i$-th bead is given by:

$$R_i^2(\gamma) \equiv \int_{t=0}^{t_f} [f_i^{\exp}(t) - f_i^{sim}(t,\gamma)]^2 dt, \tag{15}$$

where $f_i^{\exp}(t)$ is the experimental force history, and $f_i^{sim}(t,\gamma)$ is the numerical force profile derived from the discrete particle model §III, given an arbitrary chain dissipation coefficient $\gamma$. The time $t$ is counted from the striker impact moment to the time $t_f$ when the solitary wave reached the end of the chain. The total square of the residual for the entire chain is the sum of the residuals for each bead: $R^2(\gamma) \equiv \sum_{i=1}^{N} R_i^2(\gamma)$. The least square fitting is obtained by minimizing the total residual with respect to $\gamma$, yielding the condition $\partial_\gamma R^2 = 0$. For the given configuration of the dissipative granular chain and striker impact setup, this fitting gives an estimator for the chain dissipation of $\gamma|_c = -4.582$. The numerical results based on this coefficient agree well with the experimental results [Fig. 10].

To evaluate the energy losses at the wall interface, we performed a single ball impact test on various materials of bounding media (38.1-mm diameter and 76.2-mm height). Mathematically the equation of motion of a single spherical impactor on the wall can be expressed as:

$$m\ddot{u} = -A|_w u^{3/2} - \gamma|_w \dot{u} + F, \tag{16}$$

where $u$ is the approach of the spherical impactor to the bounding wall. We numerically solved this equation to find $\gamma|_w$ that matches the restitution coefficient value acquired from experiments. For the measurement of the restitution coefficients, we recorded the incident and reflected velocities using the high speed camera at the sampling frequency of 30 kHz. The measured restitution coefficients and calculated wall damping coefficients are listed in Table 2 over various materials tested.

The linear damping model certainly has limitations in capturing the complicated aspects of the solitary wave attenuation in the chain and at the wall interface. In particular we found that this model reveals noticeable errors in predicting the delay in the formation of the secondary solitary waves due to its incapability to account for the viscoelastic effects. Nonetheless this model successfully encapsulates the attenuation phenomena in the chain and at the wall, calculating with high accuracy the *TOF* of the primary solitary waves and the force amplitude ratios of the incident to the reflected waves.

Table 2. Restitution coefficients for various metallic and polymer materials and their corresponding wall dissipation coefficients derived from experimental measurements.

| Material | Restitution coefficient | Damping coefficient [kg/s] |
|---|---|---|
| Stainless steel AISI type 440C | 0.817 | -34.05 |
| Copper | 0.891 | -17.61 |
| Brass (360) | 0.865 | -21.36 |
| Aluminum 6061-T6 | 0.875 | -17.59 |
| Nylon (Polyamide) | 0.656 | -23.89 |
| Acrylic (Polymethylmethacrylate) | 0.694 | -18.13 |
| Polycarbonate | 0.884 | -5.761 |
| PTFE (Polytetrafluoroethylene) | 0.602 | -16.51 |

ACKNOWLEDGEMENTS

The authors would like to acknowledge the National Science Foundation (Grants number 0844540-CAREER and 0825345) for funding this research. The authors thank Paul Anzel for helpful suggestions.

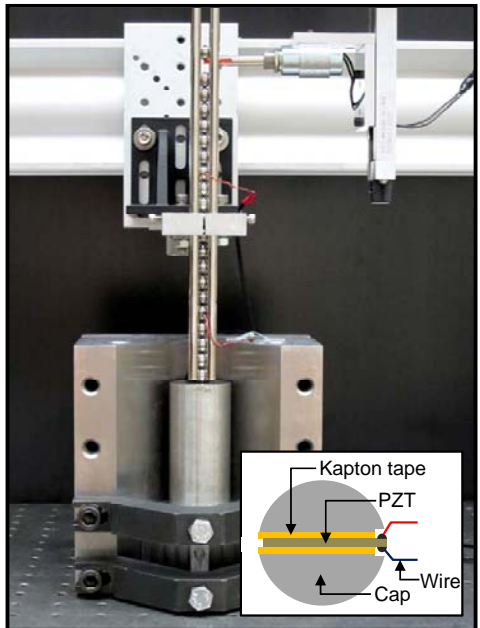

Figure 1. (Color online) Experimental setup consisting of a 20-particle granular chain vertically positioned on the top of a steel cylinder (the linear medium). The inset figure shows the schematic drawing of the instrumented particle, equipped with a piezo-ceramic layer.

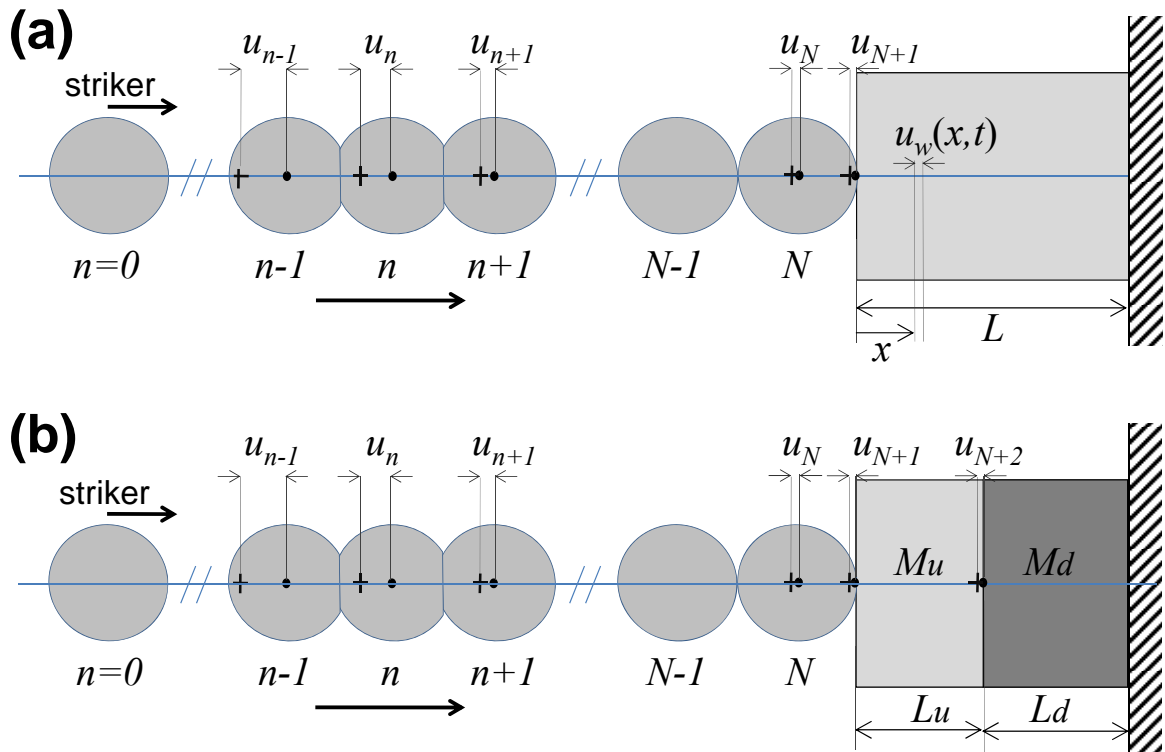

Figure 2. Schematic diagram showing the 1D chain of spherical elements in contact with (a) a uniform linear medium and (b) a composite linear medium.

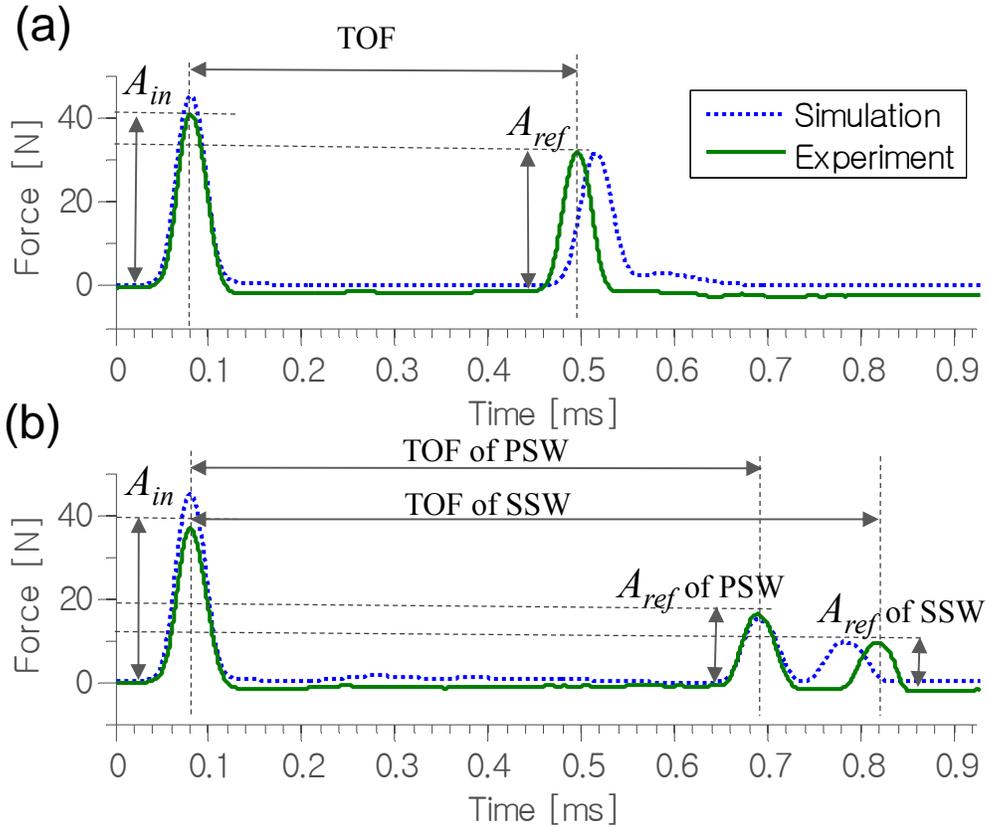

Figure 3. (Color online) Solitary wave propagation measured from the 7th bead in the chain against (a) stainless steel wall and (b) PTFE wall. The time of flight (*TOF*) represents the delay time between the incident and the reflected waves. The amplitude ratio ($A_{ref}/A_{in}$) denotes the relative magnitude of reflected solitary waves with respect to the incident solitary waves.

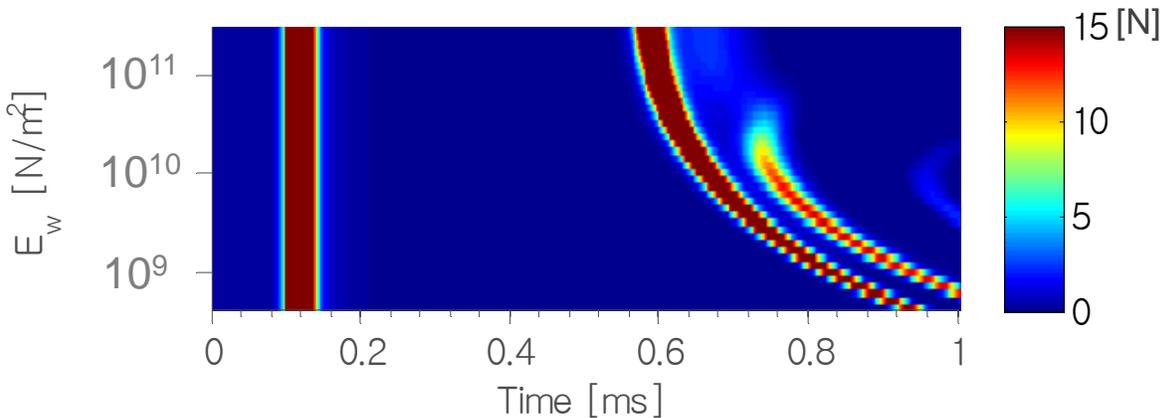

Figure 4. (Color online) Surface plot obtained from numerical simulations showing the formation of primary and secondary solitary waves in the time domain. The Y-axis reports a set of different values of elastic moduli of the linear media adjacent to the chain of spheres. Here, the first vertical line evident at ~120 μs from the impact (Time = 0) represents the arrival of the incoming solitary wave. The generation of a reflected SSW is noticeable after a critical value of elastic modulus of the contact. These simulation results are based on the force profile obtained from our numerical model for the particle number 7, and the color bar on the right denotes the amplitude of the force profiles in newtons.

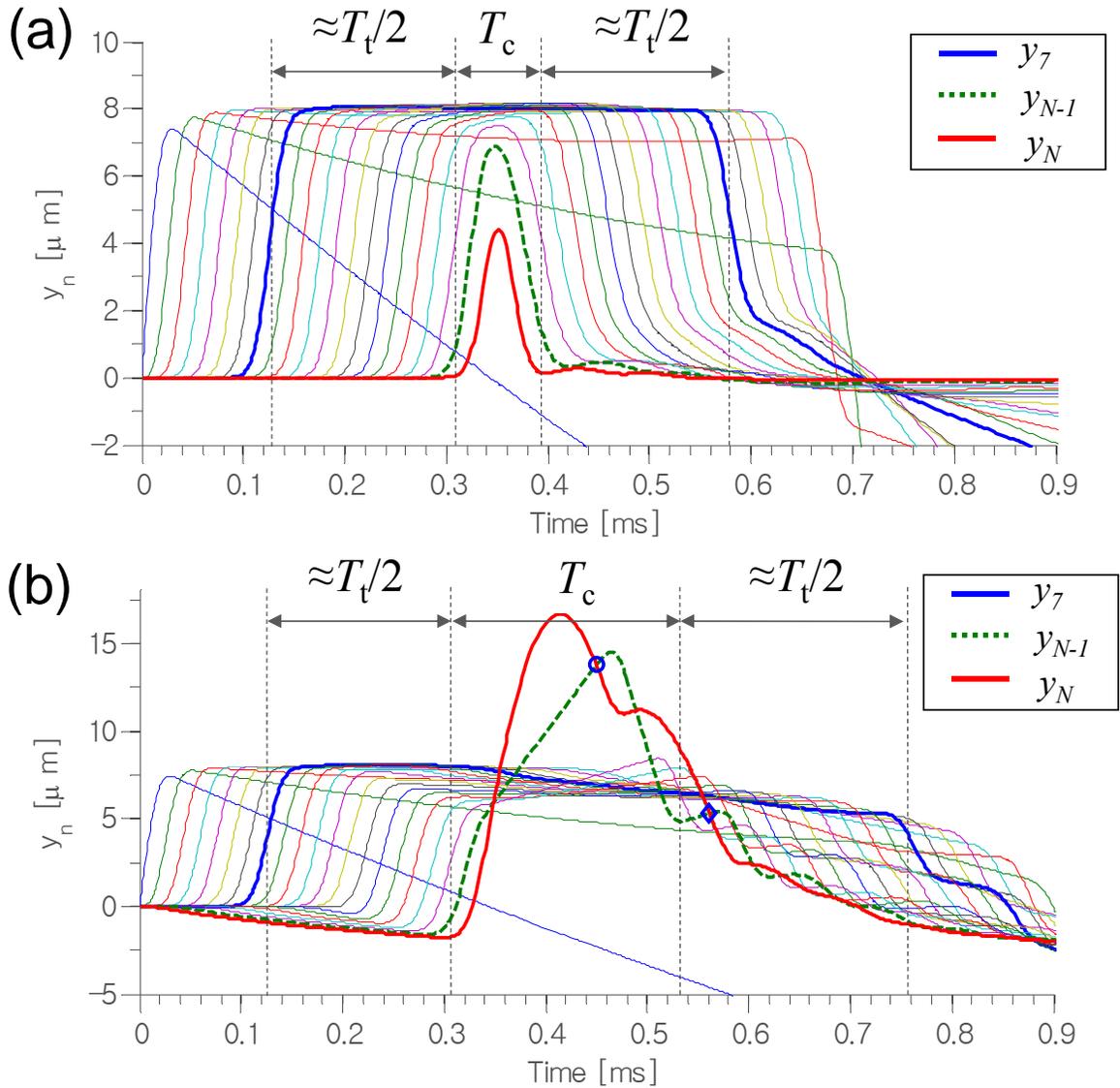

Figure 5. (Color online) Numerical results showing the displacement profiles of the striker bead (first curve on the left) and 20 particles composing the chain. (a) The stainless steel wall induces small displacement (4.39 μm) and short contact time (85 μs) of the last bead against the bounding wall (bold red line). (b) the PTFE wall allows a larger displacement of the last bead (16.66 μm) and as a result, a longer contact time (241 μs) spent on rebounding. Multiple impacts between the last and its neighboring beads are observed in the PTFE case; The first and second collisions occur at 450-μs and 561-μs points as marked in circle and square, respectively.

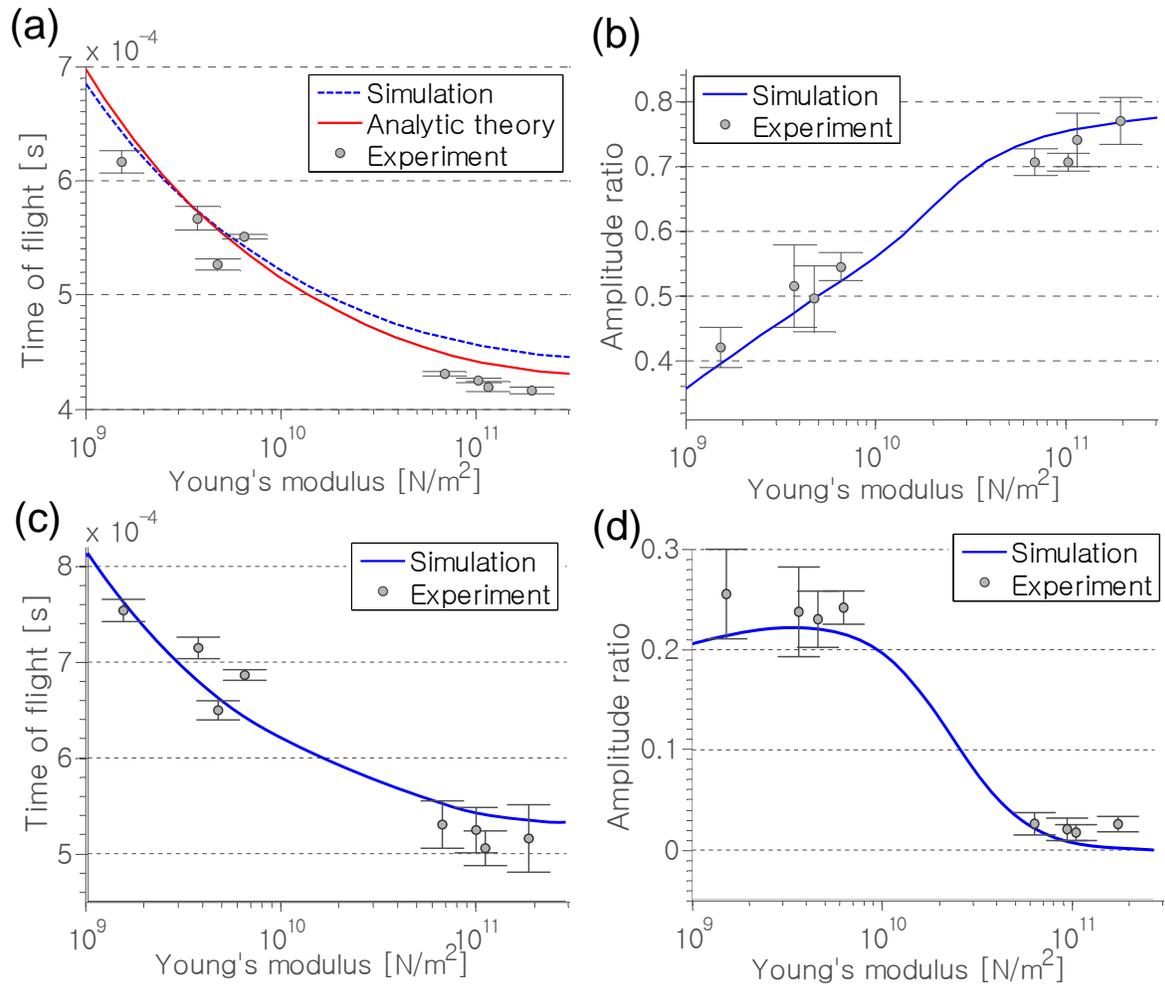

Figure 6. (Color online) Comparison of experimental, theoretical, and numerical data for the time of flight (*TOF*) and amplitude ratio of the primary and secondary reflected solitary waves as a function of the Young's modulus of the neighboring media. (a) *TOF* of the PSWs as obtained by theoretical models (solid red line), numerical calculations (dashed blue line) and experiments. (b) Amplitude ratio of the PSWs to the incident solitary wave. (c) *TOF* of the SSWs. (d) Reflection ratio of the SSWs.

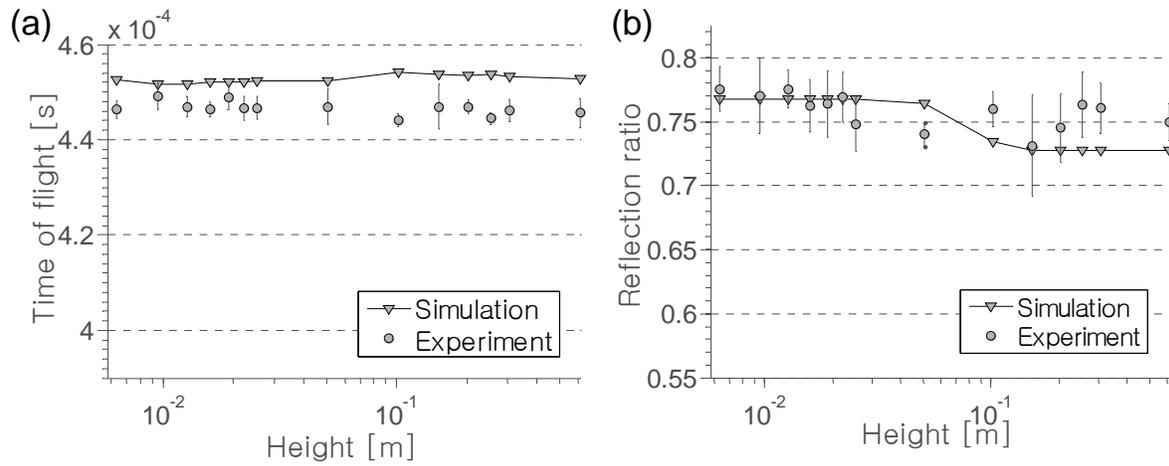

Figure 7. Time of flight and amplitude ratio of the PSWs reflected from the stainless steel slender cylinders as a function of the cylinders' heights. Numerical and experimental data are compared in the magnified Y-axis scale. (a) *TOF* of the PSW. (b) Amplitude ratio of the PSWs.

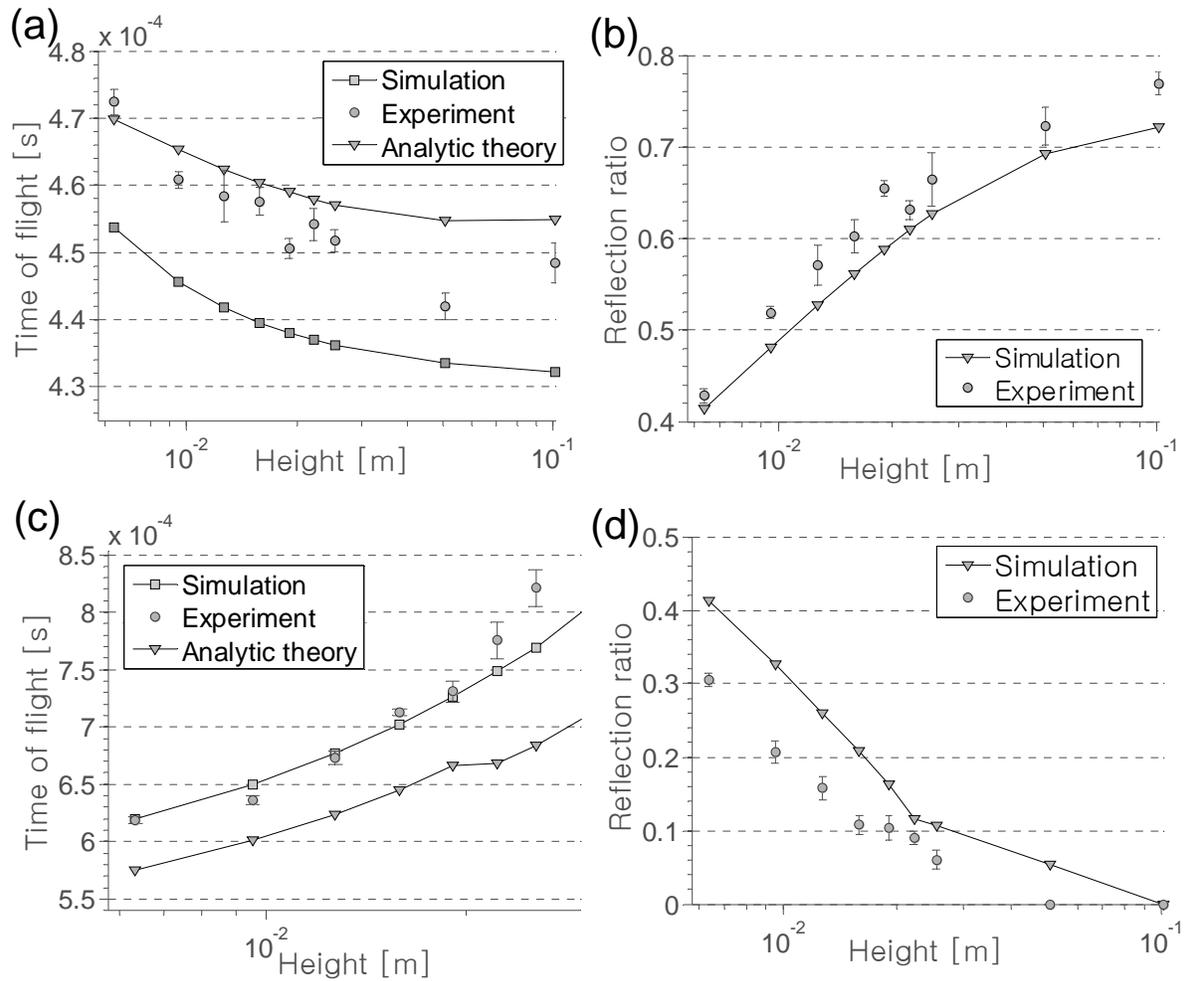

Figure 8. Comparison of experimental, theoretical, and numerical data for the time of flight and amplitude ratio of primary and secondary solitary waves, as a function of the upper layer thickness ($L_u$) in the composite media. (a) *TOF* for the PSWs. (b) Amplitude ratio for the PSWs. (c) Time of flight for the SSWs. (d) Amplitude ratio of the SSWs.

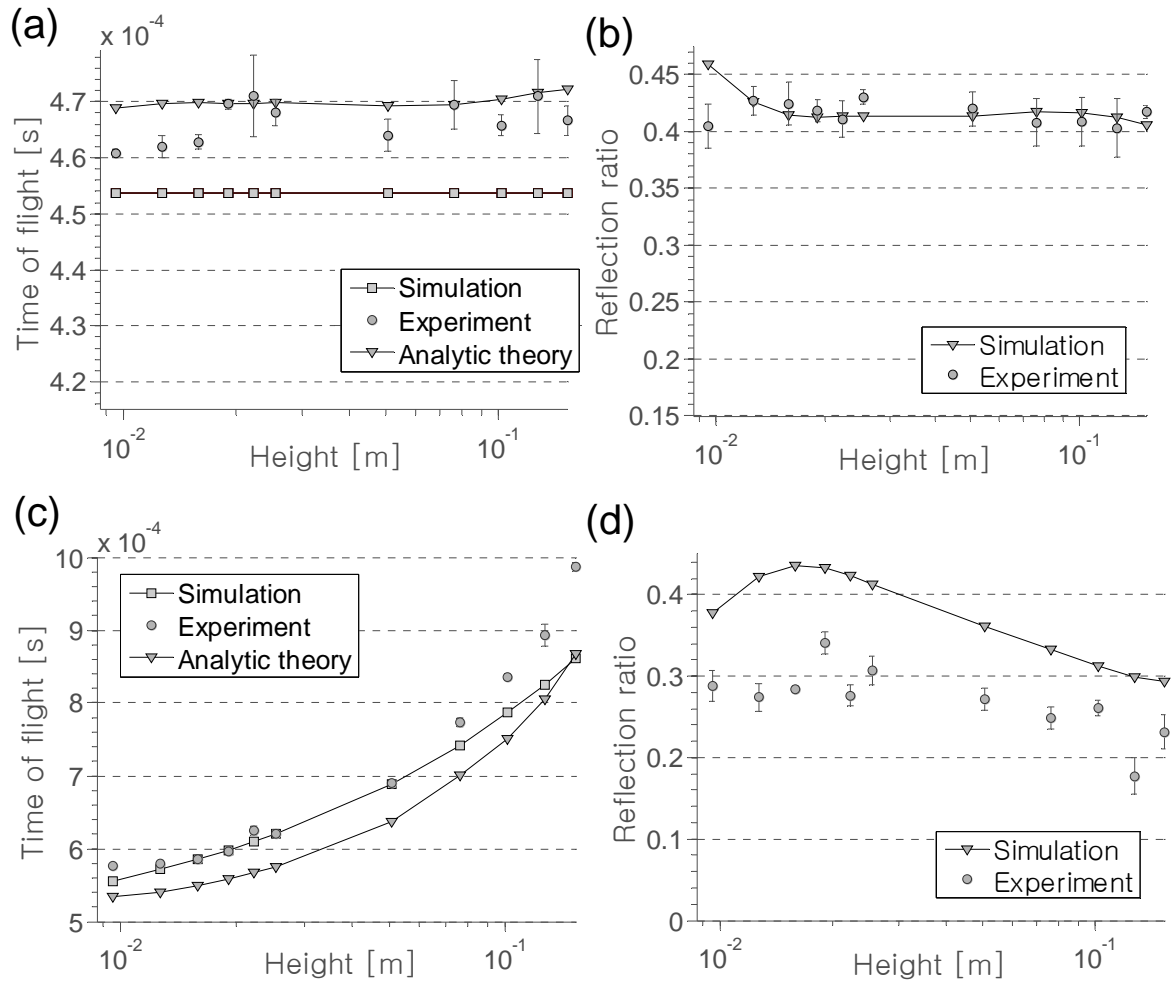
Figure 9. Time of flight and reflection ratio of the primary and secondary reflected solitary waves after the incoming solitary wave interacts with a composite medium with variable lower layer thickness. Numerical and experimental results are represented as a function of lower medium height ($L_d$). (a) *TOF* of the primary solitary waves. (b) Amplitude ratios of the PSWs. (c) Time of flight of the SSWs. (d) Amplitude ratios of the SSWs.

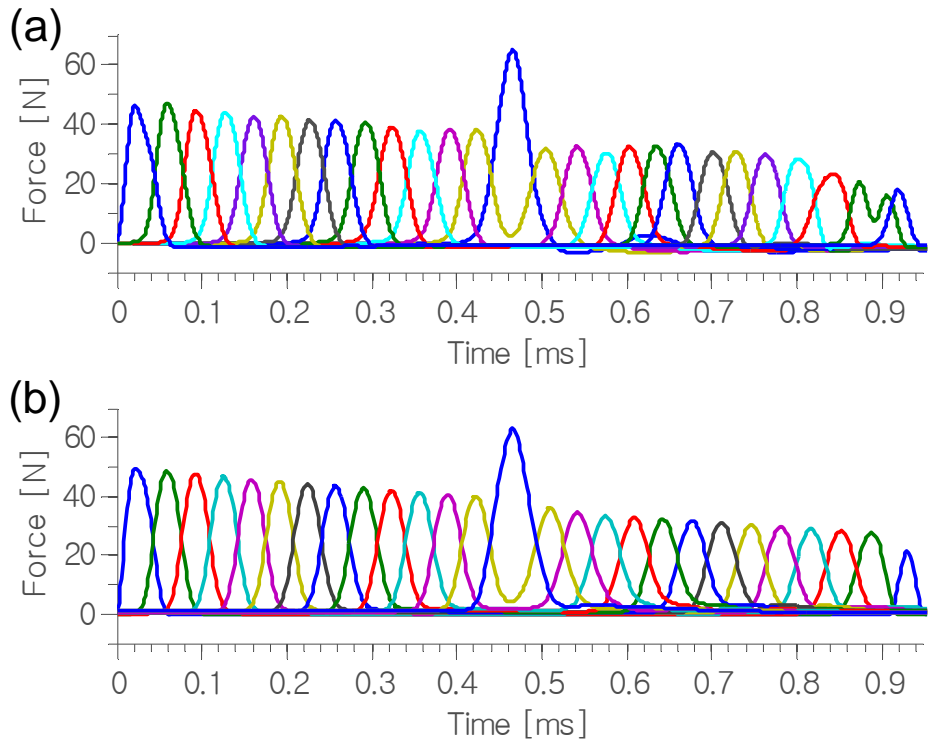

Figure 10. (Color online) Comparison of experimental and numerical force profiles of solitary waves, collected from the particles at the even-numbered positions. The spike in the center is the force measurement from the end-particle. The force profiles prior to this spike represent the solitary wave propagation before the wall reflection, whereas the latter profiles correspond to the reflected solitary waves from the stainless steel bounding wall. (a) Experimental force profiles. (b) Numerical force profiles.